%% file: torsion.tex
\documentclass[sn-mathphys]{sn-jnl}

\pdfoutput=1 

\usepackage{chngcntr}
\usepackage{framed}
\usepackage{pifont}
\usepackage{xspace}
\usepackage{relsize}

\usepackage{lmodern}
\usepackage[english]{babel}
\usepackage{upgreek}
\usepackage{times}

\usepackage{blindtext}
\usepackage{rotating}
\usepackage{subcaption}
\usepackage{sidecap}
\usepackage{soul}
\usepackage{booktabs}
\usepackage{ulem}

\usepackage{stmaryrd}
\usepackage{braket}
\usepackage{slashed}
\usepackage{multirow}
\usepackage{bm}
\usepackage{mathbbol}
\usepackage{mathrsfs}
\usepackage{empheq}
\usepackage{needspace}
\usepackage{nicefrac}
\usepackage{comment}
\usepackage{url}
\usepackage{setspace}

\usepackage{graphicx}
\usepackage[T1]{fontenc}
\usepackage{amsthm}
\usepackage{amsfonts}
\usepackage{amssymb}
\usepackage{amsmath}
\usepackage{tensor,mathtools}
\usepackage{dcolumn}
\usepackage{physics}
\usepackage{enumerate}

\usepackage{siunitx} 
\DeclareSIUnit\year{yr}
\DeclareSIUnit \parsec {pc}

\jyear{2021}%

\newcommand{\rsec}[1]{Section~\ref{#1}}

\newcommand{\req}[1]{Eq.~(\ref{#1})}
\newcommand{\rfig}[1]{Fig.~\ref{#1}}
\newcommand{\rtab}[1]{Tab.~\ref{#1}}
\renewcommand{\eqref}[1]{(\ref{#1})}
\newcommand{\eref}[1]{Eq.~\eqref{#1}}

\newcommand{\fref}[1]{Fig.~\ref{#1}}
\newcommand{\aap}{}

\newcommand{\Lag}{L}

\newcommand{\Hf}{H}

\newcommand{\LCd}{\mathcal{\Lag}}

\newcommand{\HCd}{\mathcal{\Hf}}

\renewcommand{\d}{\mathrm{d}}

\newcommand{\ho}{\omega}

\newcommand{\hoc}{\tilde{q}}

\newcommand{\onehalf}{{\textstyle\frac{1}{2}}}

\newcommand{\quarter}{{\textstyle\frac{1}{4}}}

\newcommand{\pfrac}[2]{\frac{\partial{#1}}{\partial{#2}}}

\begin{document}

\title[Article Title]{Torsion driving cosmic expansion}

\author*[1]{\fnm{Johannes} \sur{Kirsch}}\email{jkirsch@fias.uni-frankfurt.de}

\author[1]{\fnm{David} \sur{Vasak}}\email{vasak@fias.uni-frankfurt.de}
\equalcont{These authors contributed equally to this work.}

\author[1]{\fnm{Armin} \sur{van de Venn}}\email{venn@fias.uni-frankfurt.de}
\equalcont{These authors contributed equally to this work.}

\author[1,2]{\fnm{J\"urgen} \sur{Struckmeier}}\email{struckmeier@fias.uni-frankfurt.de}
\equalcont{These authors contributed equally to this work.}

\affil*[1]{\orgname{Frankfurt Institute for Advanced Studies}, \orgaddress{\street{Ruth-Moufang-Strasse~1}, \city{Frankfurt am Main}, \postcode{60438}, \state{Germany}}}

\affil[2]{\orgname{Goethe-Universit\"at}, \orgaddress{\street{Max-von-Laue-Strasse~1}, \city{Frankfurt am Main}, \postcode{60438}, \state{Germany}}}

\abstract{We study a cosmological model based on the canonical Hamiltonian transformation theory. Using a linear-quadratic approach for the free gravitational De Donder-Weyl Hamiltonian $H_\mathrm{Gr}$, the model contains terms describing a deformation of an AdS spacetime and a fully anti-symmetric torsion in addition to Einstein's theory.
The resulting extension of the Einstein-Cartan theory depends on two initially unknown constants, $\Omega_\mathrm{g}$ and $\Omega_\mathrm{s}$.
Given an appropriate choice of these parameters resulting from the analysis of asymptotics, numerical calculations were performed with $\Omega_\mathrm{\Lambda} = 0$. Values from the Planck Collaboration \cite{planck15} were used for all other required cosmological parameters. In this way, it is shown that torsion can explain phenomena commonly attributed to dark energy, and thus can replace Einstein's cosmological constant.}

\keywords{Modified theories of gravity, gauge field theory, torsion, cosmological constant}

\maketitle

\section{Introduction} \label{sec:intro}
The physical nature of dark energy is still a mystery, manifested best in the so called ``cosmological constant problem'', a 120 orders-of-magnitude mismatch between its observed and calculated values.
In fact, the concept of dark energy has, together with the yet similarly mysterious dark matter,  been an ad-hoc remedy for aligning the 
model of the Universe, based on Einstein's General Relativity (GR), with cosmological observations.

Attempts to explain both ``dark'' concepts by modifying the standard model of particles has not been validated in any of the many and extremely costly experiments carried out in the past, and little hope is manifest for changing this in the near future. An alternative and attractive research avenue for explaining the gap between theory and observations relies on  modifications of General Relativity. The numerous, so called extended theories of gravity, have delivered a variety of models linking dark energy to advanced geometrical features of spacetime, be it additional gravitational (scalar) fields, effects of torsion, or non-metricity \cite{tsamparlis81,yoon99,Medina2018,Kranas2019,unger2019big,Milton:2020tbl,Benisty2019}. The large majority of those extended theories relies, though, on ad-hoc model assumptions, often inconsistent with other observations, lacking  physical justification or even being mathematically questionable.

The objective of this work is to complement earlier analyzes of the dark energy problem \cite{vasak19a, Vasak:2021sce} based on a rigorous mathematical framework, the Covariant Canonical Gauge Gravity (CCGG) \cite{struckmeier17a, struckmeier18a}.
While in those earlier approximate calculations torsion has been identified as a good candidate for a dynamical dark energy, a detailed  model of the torsion field was missing.
Here we formulate a consistent FLRW cosmology using a totally anti-symmetric model of torsion aligned with the cosmological principle \cite{Venn2022}, and we follow up on the so called "zero-energy-Universe" conjecture that was proposed already more than 100 years ago and justified within the CCGG ansatz recently \cite{Vasak:2022gps}.

The paper is organized as follows: \rsec{sec:consistency} starts with a short review of the canonical Hamiltonian transformation theory. Using a linear-quadratic approach to the free Hamiltonian, we arrive at the so-called consistency equation, which turns out to be an extension of Einstein's field equation. The condition that the source term of gravity, the energy-momentum tensor, is conserved, is implemented with the complete anti-symmetry of the torsion tensor. In  \rsec{sec:Friedman} we apply this concept to the FLRW cosmology and obtain a system of equations that, in addition to the familiar terms of the $\Lambda$CDM standard model, also contains terms arising from the quadratic Hamiltonian and the anti-symmetric torsion. Three new parameters are also associated with these additional terms: the parameter $\Omega_\mathrm{g}$, which describes the deformation of spacetime compared to the de Sitter spacetime, the parameter $\Omega_s$, that couples the torsion to the Hamiltonian, and $s_1$, the value of torsion at the present time. We choose example values for these parameters in \rsec{sec:result} and show the corresponding results obtained by numerical solution of the system of equations. We conclude the paper with a brief comment on the cosmological implications of torsion and its ability to serve as an explanation for the expansion of the Universe.

\section{The theoretical concept} \label{sec:consistency}

\subsection{Gravity from the covariant canonical transformation theory}

The application of the canonical Hamiltonian transformation theory to semi-classical relativistic matter fields has been pioneered by Struckmeier et al. and proven to derive the Yang-Mills gauge theory from first principles \cite{struckmeier08, struckmeier13}.
At the heart of this framework is the requirement that the system dynamics is given by an action integral that remains invariant under prescribed local transformations of the original (matter) fields.
Those transformations are implemented in the covariant De Donder-Weyl (DW) Hamiltonian formalism \cite{dedonder30} by the choice of a generating function, specifically designed for any given underlying symmetry group.
That formalism unambiguously introduces symmetry dependent gauge fields and fixes their interaction with the original matter fields. The kinetic portion of the newly introduced gauge fields, here the Hamiltonian of non-interacting gravity, is not entirely determined by the gauging process, though. It is rather introduced as an educated guess based on physical considerations and empirical insights.

Applying the above framework to the diffeomorphism group paves a novel path to implementing Einstein's Principle of General Relativity to arbitrary classical relativistic systems of matter fields.
In the resulting first-order theory the spacetime geometry is described by both, the ``Lorentz'' (or ``spin'') connection $\omega\indices{^i_{j \alpha}}$ and the vierbein (tetrad) field $e\indices{^i_{\alpha}}$, where we use the convention of Misner et al. \cite{misner} that Greek indices refer to the coordinate frame, and Latin indices refer to the Lorentz (inertial) frame.
Likewise, the natural units $\hbar = c = 1$ and the metric signature $(1,-1,-1,-1)$ apply. The fundamental fields describing the dynamics of gravitation encompass, in addition to the spin connection and vierbeins, also their canonical momentum fields, the 
tensors $q\indices{_i^{\alpha\beta\xi}}$ and $k\indices{_i^{\alpha\xi}}$, respectively.

We reproduce in this chapter only the equations relevant to our objective renouncing derivations or proofs, and refer to the detailed presentation in Ref.~\cite{struckvas21}. First we note that the metric tensor $g_{\mu\nu}$ can be expressed by the vierbeins via
\begin{equation}\label{eq:g_mn}
	g_{\mu\nu}(x)=e\indices{^i_\mu}(x)\,e\indices{^j_\nu}(x)\,\eta_{ij} \, ,
\end{equation}
where $\eta_{ij}$ denotes the Minkowski metric, and the affine connection $\Gamma\indices{^{\lambda}_{\mu\nu}}$ by
\begin{equation}\label{eq:gam_mn}
	\Gamma\indices{^{\lambda}_{\mu\nu}}=-e\indices{_\mu^i}\, \left(\pfrac{e\indices{_{i}^{\lambda}}}{x^{\nu}}-\omega\indices{^k_{i \nu}} \,e\indices{_k^\lambda}\right) \, .
\end{equation}
We further assume that the spin connection $\omega\indices{^i_{j \alpha}}$ is anti-symmetric in the Lorentz indices, $\omega\indices{_{(ij) \alpha}} = 0$, which ensures metric compatibility.
The DW Hamiltonian (scalar) density, $\tilde{\HCd}_{\mathrm{Gr}}$, of spacetime dynamics extends the linear Einstein-Hilbert ansatz by  quadratic terms built from the momentum fields endowing spacetime with kinetic energy and thus inertia, and fundamentally modifying its dynamics.
In order to comply with the key observations that already gave credibility to Einstein's equation~\cite{struckmeier17a}, we set
\begin{equation} \label{CCGE}
	\tilde{\HCd}_{\mathrm{}} =
	\tilde{\HCd}_{\mathrm{Gr}}  + \tilde{\HCd}_{\mathrm{matter}}
\end{equation}
with a quadratic-linear ansatz $\tilde{\HCd}_{\mathrm{Gr}} = \tilde{\HCd}_{\mathrm{Gr}}(\tilde{q}^2 e^4, \, \tilde{q} e^2, \, \tilde{k}^2 e^4, \, \varepsilon)$ which in full detail reads
\begin{align}
	\tilde{\HCd}_{\mathrm{Gr}}&=\frac{1}{4g_1\varepsilon}\tilde{q}\indices{_l^{m\alpha\beta}}
	\tilde{q}\indices{_m^{l\xi\lambda}}\,\eta_{kn}\,\eta_{ij}\,e\indices{^k_\alpha}\,e\indices{^n_\xi}\,e\indices{^i_\beta}\,e\indices{^j_\lambda} -g_2\,\tilde{q}\indices{_l^{m\alpha\beta}}e\indices{^l_\alpha}\,e\indices{^n_\beta}\,\eta_{mn}\nonumber\\
	&\quad+\frac{1}{2g_3\varepsilon}\tilde{k}\indices{_l^{\alpha\beta}}\tilde{k}\indices{_m^{\xi\lambda}}\,
	\eta^{lm}\eta_{kn}\,\eta_{ij}\,e\indices{^k_\alpha}\,e\indices{^n_\xi}\,e\indices{^i_\beta}\,e\indices{^j_\lambda} +g_4\varepsilon \, .
	\label{def:Hdyn}
\end{align}
The tilde is used to denote tensor densities and $\varepsilon := \det (e\indices{_\mu ^i}) = \sqrt{-g_{\mu\nu} \, }$. $\tilde{\HCd}_{\mathrm{matter}}$ involves the coupling of matter fields to curved spacetime.
The coupling constants $g_1$, $g_2$, $g_3$, and $g_4$ have dimensions $[g_1] = 1$, $[g_2] = L^{-2}$, $[g_3] = L^{-2}$, and $[g_4] = L^{-4}$. $g_4$ is usually identified with the vacuum energy density of matter and leads to the so-called cosmological constant problem \cite{weinberg89} of General Relativity.

The gauging process results in the action integral
\begin{align}
	S=\int_{V}\d^4x \, \tilde{\LCd}= &\int_{V}\d^4x\Bigg\{\onehalf\tilde{k}\indices{_i^{\mu\nu}}\left(\pfrac{e\indices{^i_\mu}}{x^\nu}-\pfrac{e\indices{^i_\nu}}{x^\mu}
	+\ho\indices{^i_{j\nu}}\,e\indices{^j_\mu}-\ho\indices{^i_{j\mu}}\,e\indices{^j_\nu}\right)\nonumber\\
	&+\onehalf\hoc\indices{_i^{j\mu\nu}}\left(\pfrac{\ho\indices{^i_{j\mu}}}{x^\nu}
	-\pfrac{\ho\indices{^i_{j\nu}}}{x^\mu}+\ho\indices{^i_{n\nu}}\,\ho\indices{^n_{j\mu}}
	-\ho\indices{^i_{n\mu}}\,\ho\indices{^n_{j\nu}}\right) \nonumber \\ &-\tilde{\HCd}_{\mathrm{Gr}}+\tilde{\LCd}_{\mathrm{matter}}\Bigg\},\label{actionintegral2a}
\end{align}
where the total Lagrangian, a world scalar density, is split up into the modified gravity Lagrangian, displayed explicitly as a Legendre transform of the Hamiltonian $ \tilde{\HCd}_{\mathrm{Gr}}$ of \eref{def:Hdyn}, and the yet unspecified~$\tilde{\LCd}_{\mathrm{matter}}$.
Variation of \eref{actionintegral2a} with respect to $\hoc\indices{_i^{j\mu\nu}}$ gives the first canonical equation,
\begin{equation}\label{eq:omega-deri}
	\pfrac{\tilde{\HCd}_\mathrm{Gr}}{\hoc\indices{_i^{j\mu\nu}}}=
	\onehalf\left( \pfrac{\ho\indices{^i_{j\mu}}}{x^\nu}-\pfrac{\ho\indices{^i_{j\nu}}}{x^\mu}+\ho\indices{^i_{n\nu}}\,\ho\indices{^n_{j\mu}}-\ho\indices{^i_{n\mu}}\,\ho\indices{^n_{j\nu}}\right) =: \onehalf R\indices{^i_{j\nu\mu}}\, .
\end{equation}
It is straightforward to prove with \req{eq:gam_mn} that $R\indices{^i_{j\nu\mu}}$ is equivalent to the Riemann-Cartan curvature tensor \footnote{Note that assuming in addition to \req{eq:gam_mn} the anti-symmetry of the spin connection ensures metric compatibility $g_{\alpha\beta;\nu} =0$, to which the following investigation is restricted.}
\begin{equation}
	R\indices{^{i}_{j\mu\nu}} \, e\indices{_i^\eta}\,e\indices{^j_\alpha} \equiv
	R\indices{^{\eta}_{\alpha\mu\nu}} \, =
	\, \pfrac{\Gamma\indices{^{\eta}_{\alpha\nu}}}{x^{\mu}} -
	\, \pfrac{\Gamma\indices{^{\eta}_{\alpha\mu}}}{x^{\nu}} +
	\, \Gamma\indices{^{\eta}_{\xi\mu}}\Gamma\indices{^{\xi}_{\alpha\nu}} -
	\, \Gamma\indices{^{\eta}_{\xi\nu}}\Gamma\indices{^{\xi}_{\alpha\mu}} \, .
\end{equation}
The canonical equation \eqref{eq:omega-deri} with the specific ansatz \eqref{def:Hdyn} for $\tilde{\HCd}_\mathrm{Gr}$ fixes the relation of $q\indices{^i_{j\mu\nu}}$ to $R\indices{^i_{j\mu\nu}}$.
Written in the coordinate frame this gives
\begin{equation} \label{eq:R-q}
	q\indices{_{\xi\lambda\mu\nu}}=g_1 \, \left( R\indices{_{\xi\lambda\mu\nu}} -\bar{R}\indices{_{\xi\lambda\mu\nu}} \right)  \, ,
\end{equation}
where
\begin{equation} \label{def:maxsymR}
	\bar{R}\indices{_{\xi\lambda\mu\nu}}=g_2\left(g_{\xi\mu}\,g_{\lambda\nu}-g_{\xi\nu}\,g_{\lambda\mu}\right)
\end{equation}
is the Riemann curvature tensor of the maximally symmetric (Anti) de Sitter spacetime with the Ricci scalar curvature $12\,g_2$.
The affine momentum $q\indices{_{\xi\lambda\mu\nu}}$ thus accounts for deformations of the geometry relative to the (A)dS ground state.

So far, no assumptions regarding the symmetry of the affine connection \eqref{eq:gam_mn} has been made. In the following we retain torsion as an additional structural element of the underlying geometry. Such a (Riemann-Cartan) manifold extends the affine connection from the Christoffel symbol of the Einstein-Hilbert ansatz to
\begin{equation}\label{eq:gammaing}
	\Gamma\indices{^{\lambda}_{\mu\nu}} =
	\genfrac{\lbrace}{\rbrace}{0pt}{0}{\lambda}{\mu\nu}+
	K\indices{^{\lambda}_{\mu\nu}} \, ,
\end{equation}
where the contortion tensor
\begin{equation} \label{def:contortiona}
	K\indices{_{\lambda}_{\mu\nu}} \,=\,
	S\indices{_{\lambda\mu\nu}} - S\indices{_{\mu\lambda\nu}} + S\indices{_{\nu\mu\lambda}}\,=\,-K\indices{_{\mu}_{\lambda\nu}}
\end{equation}
is built from Cartan's torsion tensor
\begin{equation} \label{def:torsiontensor}
	S\indices{^\lambda_{\mu\nu}} = \onehalf(\Gamma\indices{^{\lambda}_{\mu\nu}}- \Gamma\indices{^{\lambda}_{\nu\mu}}) \, .
\end{equation}
Taking into account the relation
\begin{equation}
	S\indices{^i_{\mu\nu}} := \pfrac{e\indices{^i_\mu}}{x^\nu}-\pfrac{e\indices{^i_\nu}}{x^\mu}
	+\ho\indices{^i_{j\nu}}\,e\indices{^j_\mu}-\ho\indices{^i_{j\mu}}\,e\indices{^j_\nu}
	\equiv
	S\indices{^\lambda_{\mu\nu}}\, e\indices{^i_\lambda}
\end{equation}
that again follows from \req{eq:gam_mn}, the variation of the action integral with respect to the field ${\tilde{k}\indices{_i^{\mu\nu}}}$ yields the second canonical equation
\begin{align}
	\pfrac{\tilde{\HCd}_\mathrm{Gr}}{\tilde{k}\indices{_i^{\mu\nu}}} = \frac{1}{g_3}
	\,k\indices{^i_{\mu\nu}} = S\indices{^i_{\mu\nu}} \,  \label{eq:e-deri3}
\end{align}
relating the momentum ${\tilde{k}\indices{_i^{\mu\nu}}}$ to torsion.
Finally, the so-called consistency equation that extends Einstein gravity~\cite{struckmeier17a, struckmeier18a, struckvas21} is obtained from a combination of all the canonical equations including matter dynamics. It can be written as the local balance equation (also called the ``zero-energy Universe'' paradigm going back to Lorentz \cite{lorentz1916} and Levi-Civita \cite{levi-civita1917} and re-derived in~\cite{Vasak:2022gps} from CCGG)
\begin{equation} \label{eq:modEinstein}
	\Theta\indices{_\mu^\nu} + T\indices{_\mu^\nu} = 0
\end{equation}
with
\begin{subequations}
	\begin{alignat}{4}
		\Theta\indices{_\mu^\nu} &:= \frac{1}{\sqrt{-g}}\,e\indices{^i _\mu}\pfrac{\tilde{\HCd}_{\mathrm{Gr}}}{e\indices{^i_{\nu}}} \label{def:Q0} \\
		T\indices{_\mu^\nu} &:= \frac{1}{\sqrt{-g}}\,e\indices{^i _\mu}\pfrac{\tilde{\HCd}_{\mathrm{matter}}}{e\indices{^i_{\nu}}} \, .
	\end{alignat}
\end{subequations}
and is similar to the stress-strain relation in elastic media. In analogy to the energy-momentum (``stress-energy'') tensor of matter, $T\indices{^\mu^\nu}$, we interpret $\Theta\indices{^\mu^\nu}$ as the energy-momentum (``strain-energy'') tensor of spacetime \footnote{This also implies that the total energy of the Universe is zero, consistent with Jordan's conjecture, cf. \cite{jordan39}. Taking the vacuum expectation value of the ``quantum analogue'' of this equation, we would expect the vacuum energy densities of gravity and the matter to cancel each other, perhaps up to some residual value that can be identified with $g_4 \approx 0$.}.  Calculating now the strain-energy tensor \eqref{def:Q0} with the Hamiltonian~\eqref{def:Hdyn}, and substituting \req{eq:R-q} for the momentum tensor, gives
\begin{equation} \label{def:Theta2}
	\begin{split}
	\Theta\indices{^\mu^\nu} = &-g_1 \, Q\indices{^\mu^\nu} + 2g_1 g_2\, \left(G\indices{^\mu^\nu} + 3g_2 \, g\indices{^\mu^\nu}\right) \\&+2 g_3 \left( S^{\xi \alpha \mu} S\indices{_{\xi \alpha}^\nu} - \onehalf S^{\mu \alpha \beta} S\indices{^\nu _{\alpha \beta}}- \quarter g^{\mu \nu} S_{\xi \alpha \beta} S^{\xi \alpha \beta}\right) +g_4 \, g\indices{^\mu^\nu}
	\end{split}
\end{equation}
where
\begin{equation}
	G\indices{^\mu^\nu} := R\indices{^{(\mu\nu)}}
	- \onehalf g\indices{^\mu^\nu} \, R
\end{equation}
is the Einstein tensor \footnote{The Einstein tensor as derived from the canonical equations of motion contains only the symmetrized Ricci tensor. While in the absence of torsion the Ricci tensor is symmetric, it is not the case for non-zero torsion. The anti-symmetric Ricci tensor interacts with the spin density of matters.}, and
\begin{equation} \label{eq:Q2}
	Q\indices{^\mu^\nu} :=  R^{\alpha\beta\gamma\mu}\, R\indices{_{\alpha\beta\gamma}^{\nu}}
	- \quarter g\indices{^\mu^\nu} \, R^{\alpha\beta\gamma\xi}\, R_{\alpha\beta\gamma\xi},
\end{equation}
is a trace-free, (symmetric) quadratic Riemann-Cartan concomitant.  \eref{def:Theta2} is a generalization of the l.h.s. of the Einstein equation in three aspects. Firstly, a Palatini equivalent formalism is used, i.e. the spin connection and the vierbeins are independent fields, torsion of spacetime is admitted, and a quadratic Riemann-Cartan term is added. In the Lagrangian constructed by Legendre transformation that term is built from the Kretschmann scalar $R^{\alpha\beta\mu\nu}\, R_{\alpha\beta\mu\nu}$.
Combining equations \eqref{def:Theta2}-\eqref{eq:Q2} the consistency equation then reads:
\begin{align} \label{eq:consist}
	\begin{split}
		& g_1\left( R^{\alpha\beta\gamma\mu}\, R\indices{_{\alpha\beta\gamma}^\nu} \!
		- \quarter g^{\mu\nu} \, R^{\alpha\beta\gamma\xi}\, R_{\alpha\beta\gamma\xi} \right) - \frac{1}{8\pi G}\, \left( R^{(\mu \nu)}
		- \onehalf  g^{\mu\nu} R +  g^{\mu\nu} \Lambda_0 \right) \\
		& - 2 g_3 \left( S^{\xi \alpha \mu} S\indices{_{\xi \alpha}^\nu} - \onehalf S^{\mu \alpha \beta} S\indices{^\nu _{\alpha \beta}}- \quarter g^{\mu \nu} S_{\xi \alpha \beta} S^{\xi \alpha \beta}\right)  = T^{(\mu \nu)}.
	\end{split}
\end{align}
Here the coupling constants $g_2$ and $g_4$ in \eref{eq:modEinstein} have been expressed in terms of the gravitational coupling constant $G$ and a constant~$\Lambda_0$:
\begin{subequations}
	\begin{alignat}{4}
		& g_1\,g_2 \equiv \frac{1}{16\pi G} = \onehalf M_p^2 \label{def:constantsg1}\\
		& 6g_1\,g_2^2+g_4 \equiv \frac{\Lambda_0}{8\pi G} \; = M_p^2\,\Lambda_0.
		\label{def:constantsg3}
	\end{alignat}
\end{subequations}
$M_p := \sqrt{1/8\pi G}$ is the reduced Planck mass.
These relations, that can be derived from the weak gravity limit \cite{kehm17}, allow to align the above field equation with the notation of GR.
Moreover, combining both equations yields
\begin{equation} \label{def:constantsg32}
	\Lambda_0 = 3g_2+8\pi G g_4= \frac{1}{M^2_p}\left(\frac{3 M_p^2}{2 g_1}+g_4\right) \, .
\end{equation}
Obviously,  $\Lambda_0$ is not a fundamental constant like Einstein's cosmological term but it is derived as a combination of  the (A)dS curvature of the ground state of space-time and the vacuum energy of matter~\cite{vasak19a}.
The parameter~$g_1$ is the deformation parameter of the theory \footnote{This result reminds of earlier approaches under the heading of de Sitter relativity to derive the cosmological constant and to explain cosmic coincidence and time delays of extra-galactic gamma-ray flares (see for example~\cite{Aldrovandi2009}).} as it determines the relative strength of the quadratic Riemann-Cartan extension of Einstein gravity.
(The coupling constant~$g_2 = M_p^2/2g_1$ is thus proportional to the inverse of that deformation parameter.)
Setting $\Lambda_0 \equiv 0$ as follows from the zero-energy condition \eqref{eq:modEinstein} relates \cite{Venn2022,Vasak:2022gps}  then constants $g_1$ and $g_2$ to the vacuum energy density $g_4$:
\begin{equation} \label{eq:gi(g4)}
	g_1 = - \frac{3M_p^4}{2g_4}; \,\,\,\,\,  g_2 = - \frac{g_4}{3M_p^2}.
\end{equation}

\subsection{The torsion model}

Requesting the stress-energy tensor of matter, $T\indices{^{(\nu\mu)}}$, to be covariantly conserved in the CCGG theory leads in general to the necessity for adjusting the affine connection beyond the Levi-Civita relation by invoking torsion of spacetime as a new structural element of the spacetime geometry, specific to this requirement.
We show that for classical matter this can be achieved if the torsion tensor is totally anti-symmetric.
The objective here is to construct a torsion tensor such that
\begin{equation} \label{eq:covderTbar1}
	{\nabla}_\nu {T}\indices{^{(\mu\nu)}} \equiv {\nabla}_\nu \bar{T}\indices{^{(\mu\nu)}} =\bar{\nabla}_\nu \bar{T}\indices{^{(\mu\nu)}} =0
\end{equation}
holds to align with the assumption underlying standard $\Lambda$CDM cosmology by defining the scaling law for (conserved) matter and radiation, with the convention that overbared quantities are calculated with Christoffel symbol. Then also
\begin{equation} \label{eq:theta0}
	\nabla_\mu \Theta^{\nu\mu} = 0
\end{equation}
must hold.  Consider now
\begin{equation} \label{eq:covderTbar2}
	{\nabla}_\mu \bar{T}\indices{^{(\nu\mu)}} =\bar{\nabla}_\mu \bar{T}\indices{^{(\nu\mu)}}
	+ K\indices{^{\nu}_{\alpha\mu}}\, \bar{T}\indices{^{(\alpha\mu)}}
	+ K\indices{^{\mu}_{\alpha\mu}}\, \bar{T}\indices{^{(\nu\alpha)}} =0.
\end{equation}
with the contortion tensor defined in \req{def:contortiona}. By requirement the first term on the r.h.s. is zero. Due to the symmetry of the stress-energy tensor, only the symmetric portion of the contortion tensor contributes to the second term
\begin{align*}
	K\indices{_{\nu}_{(\alpha\mu)}} =
	S\indices{_{\mu}_{\alpha\nu}} + S\indices{_{\alpha}_{\mu\nu}}
	= 2 S\indices{_{(\alpha\mu)}_{\nu}},
\end{align*}
while in the third term it is its non-vanishing trace:
\begin{equation*}
	K\indices{^{\mu}_{\alpha\mu}} =
	S\indices{^{\mu}_{\alpha\mu}} - S\indices{_{\alpha}^\mu_\mu} + S\indices{_{\mu}_\alpha^\mu}
	= 2 S\indices{^{\mu}_{\alpha\mu}} =: 2 S_\alpha
\end{equation*}
Then in \eref{eq:covderTbar2} the torsion dependent terms become
\begin{equation}
	2 \left( S\indices{_{(\alpha\mu)}_{\nu}}\, \bar{T}\indices{^{(\alpha\mu)}}
	+ S^\alpha \, \bar{T}\indices{_{(\nu\alpha)}} \right) =
	2 \left( S\indices{_{(\alpha\mu)}_{\nu}}
	+ S_\alpha \, g\indices{_\nu_\mu}  \right) \bar{T}\indices{^{(\alpha\mu)}} =
	0.
\end{equation}
We observe that if $S\indices{_{\alpha\mu}_{\nu}}$ is anti-symmetric in $\mu\alpha$, giving a totally anti-symmetric torsion tensor~\footnote{For similar consideration on the symmetry of the torsion implied by the cosmological principle see \cite{tsamparlis81, yoon99}}, then $S\indices{_{(\alpha\mu)}_{\nu}} = 0$ and also $S_\mu$ vanishes. Selecting a totally anti-symmetric torsion tensor is thus a (not necessary but) sufficient condition for covariant conservation of the relevant symmetric portion of the stress-energy tensor. Then also the torsion and contortion tensors are identical, $K\indices{_{\lambda}_{\mu\nu}} \equiv
S\indices{_{\lambda\mu\nu}}$, and \req{eq:gammaing} reads now
\begin{equation}\label{eq:gammaS}
	\gamma\indices{^{\lambda}_{\mu\nu}} =
	\genfrac{\lbrace}{\rbrace}{0pt}{0}{\lambda}{\mu\nu}+
	S\indices{^{\lambda}_{\mu\nu}}
\end{equation}
Because a totally anti-symmetric rank-3 tensor in four dimensions has only four independent
elements, we can re-write the torsion tensor as
\begin{equation}
	{S}\indices{_{\alpha\mu\nu}} = \frac{1}{\sqrt{3!}}\epsilon\indices{_{\alpha\mu\nu\sigma}} \, {s}^\sigma
\end{equation}
where we use the totally anti-symmetric covariant Levi-Civita tensor density
$\epsilon\indices{_{\alpha\mu\nu\sigma}}$ that is invariant under chart transitions.
This relation can be reversed to express the axial vector density $s^\sigma$ using the contravariant tensor density $\epsilon \indices{^{\alpha \mu \nu \sigma}}$:
\begin{equation}
	{s}^\sigma = \frac{1}{\sqrt{3!}} \epsilon \indices{^{\sigma\alpha\mu\nu }} \, {S}\indices{_{\alpha\mu\nu}}.
\end{equation}
In order to preserve the cosmological principle, i.e. the homogeneity and isotropy of the maximally symmetric 3-dimensional space, we pursue a similar ansatz as done in \cite{Kranas2019} and apply for $s^\sigma$ a time-like vector density
\begin{equation} \label{eq:s0}
	s\indices{^{\sigma}} = (s_0,0,0,0)
\end{equation}
where $s_0$ is a scalar function depending only on time. Thus, the $g_3$ proportional term in \req{eq:consist} can be evaluated into
\begin{align}
	S^{\xi \alpha \mu} S\indices{_{\xi \alpha \nu}} - \onehalf S^{\mu \alpha \beta} S\indices{_{\nu \alpha \beta}}- \quarter \delta^\mu _\nu \, S_{\xi \alpha \beta} S^{\xi \alpha \beta}
	= - \left(\frac{s_0}{\sqrt{3!}}\right)^2 \,
	\left(\begin{matrix}
		\frac{3}{2} & 0 & 0 & 0 \\
		0 & \frac{1}{2} & 0 & 0 \\
		0 & 0 & \frac{1}{2} & 0 \\
		0 & 0 & 0 & \frac{1}{2}
	\end{matrix}\right) \, .
\end{align}
The concept presented in this section provides a complete description of a Riemannian, metric compatible geometry with total anti-symmetric torsion. The next section deals with its impact on the standard model of cosmology.

\section{The extended Friedman equations} \label{sec:Friedman}
Following the Cosmological Principle, we deploy the Friedman-Lemaître-Robertson-Walker (FLRW) metric where the FLRW line element in spherical co-moving coordinates ($t, r, \theta, \phi $) reads
\begin{align}
	ds^2 = dt^2-a^2(t) \left[\frac{dr^2}{1-K_0 r^2}+
	r^2(d\theta ^2+\sin^2\theta d\phi ^2) \right].
	\label{eq:FLRWmetric}
\end{align}
The parameter $K_0$ fixes the type of the underlying spatial geometry: $K_0 = 0$ flat, $K_0 > 0$ spherical, $K_0 < 0$ hyperbolic. The dimensionless scale factor $a(t)$ characterizes the relative size of space-like hypersurfaces at different times. 
$K_0$ has the dimension $L^{-2}$ \footnote{Note that we have the choice to either define the scale parameter $a$ \emph{or} the spatial curvature parameter $K$ as dimensionless, but not both at the same time. This is often ignored in the literature.} , and $R_K := K_0^{-1/2}$ is for positive $K_0$ (closed Universe) the radius of the 3D space.

The material content of the Friedman Universe is modelled by non-interacting perfect fluids made of baryonic and cold dark matter and radiation giving the diagonal stress-energy tensor
\begin{equation}
	T\indices{^\mu_\nu} = \sum_{i=r,m}\, \mathrm{diag}(\rho_i,-p_i,-p_i,-p_i).
	\label{def:perfectemtensor}
\end{equation}
The energy densities $\rho_i$ and pertinent pressures $p_i$ are functions of the global time $t$ only, and the index $i$ tallies here the two contributing components: $i=m$ for baryonic and dark matter and $i=r$ for radiation. The equation of state (EOS) for a perfect fluid is assumed~\cite{weinberg72} to have the barotropic linear form
\begin{equation} \label{EOSgeneric}
	p_i = w_i\, \rho_i \, ,
\end{equation}
here with the ``dust'' condition $w_\mathrm{m}=0$ and the relativistic particle condition $w_\mathrm{r}=\nicefrac{1}{3}$.

Applying the FLRW-metric to the consistency equation \eqref{eq:consist} with a totally anti-symmetric torsion, and inserting the stress-tensor \eqref{def:perfectemtensor}, leads to the extended Friedman equations \footnote{Many of the subsequent equations are derived or verified with the help of the software tool ``Maple'' released by Maplesoft\texttrademark, see \url{https://www.maplesoft.com/}.}
\begin{subequations}
	\begin{align}
		& - 8\pi Gg_1\left[\left(\frac{\dot{a}^2+K_0}{a}\right)^2-\ddot{a}^2 \right]+\dot{a}^2+K_0  -\frac{1}{3}\Lambda_0 a^2+B(a,s_0)  \nonumber \\
		& =
		\frac{8\pi G}{3} a^2\sum_{i=m,r} \rho_i \label{eq:Fr1} \\
		& a \ddot{a} + \dot{a}^2 +K_0 - \frac{2}{3}\Lambda_0 a^2 +(8\pi G g_3-1)\frac{s^2_0}{6} a^2 = \frac{4\pi G}{3} a^2 \rho_\mathrm{m}
	\end{align} \label{eq:Fr_ab}
\end{subequations}
where
\begin{align*}
	B(a,s_0) : = \, & 8\pi Gg_1\biggl(\frac{s^2_0}{6}(-\frac{s^2_0}{6} a^2+5\dot{a}^2+2K_0) - \frac{1}{6} (\dot{s_0}^2 a^2 +2\dot{s_0}s_0\dot{a}a) \biggr)  \\ & +(8\pi G g_3-1) \frac{s_0^2}{6} a^2 \, .
\end{align*}
For $s_0 = 0$ and thus $B = 0$ these equations reduce to the familiar Friedman equations.
Inspired by the $\Lambda$CDM model we introduce the cosmological parameters $\Omega_\mathrm{m}, \, \Omega_\mathrm{r}, \, \Omega_\Lambda \! =\Lambda_0 /3H_{0}^2$, \mbox{$\Omega_\mathrm{K} =-K_0/H_{0}^2$}, the Hubble constant \mbox{$H_0 = h \, H_{100} = h \cdot \SI{100}{\kilo\metre\per\second\per\mega\parsec}$}, and the dimensionless time $\tau = t \,H_{0}$.
Furthermore, we assume the scaling of $\rho_\mathrm{m}$ and $\rho_\mathrm{r}$ according to the $\Lambda$CDM model
\begin{align} \label{eq:rho_mr}
	\begin{split}
		& \rho_\mathrm{m}(a) = \rho_{cr} \, \Omega_\mathrm{m} \, a^{-3} \\
		& \rho_\mathrm{r}(a) = \rho_{cr} \, \Omega_\mathrm{r} \, a^{-4}
	\end{split}
\end{align}
where the critical density $\rho_{cr}$ is defined by
\begin{equation} \label{eq:rho_cr}
	\rho_{cr} = \frac{3H_{0}^2}{8\pi G} = 3M_p^2\,H_{0}^2.
\end{equation}
Then the above equations can be transformed to a set of dimensionless equations:
\begin{subequations} \label{eq:Fr12}
	\begin{align}
		& \dot{a}^2 + V(a) = \Omega_{K} \label{eq:Fr12a} \\
		& \ddot{a} a + \dot{a}^2 -2Ma^2 -\Omega_{K}+(\Omega_s-1) s^2 a^2 = 0 \label{eq:Fr12b}
	\end{align}
\end{subequations}
with the definitions
\begin{subequations} \label{eq:V}
	\begin{align}
		& M(a) := \frac{1}{4}\Omega_\mathrm{m} a^{-3} + \Omega_\mathrm{\Lambda} \\
		& \Omega_\mathrm{g} := \frac{1}{32\pi G \, H_{0}^2 \, g_1} \label{eq:def_Og} \\
		& \Omega_s := 8\pi G  g_3 \label{eq:def_Os} \\
		& s := \frac{s_0}{\sqrt{3!}H_{0}} \\
		& V(a) := V_0(a)+V_\mathrm{{geo}}(a)+V_\mathrm{{tor}}(a,s) \\
		& V_0(a) := -\Omega_\mathrm{m} a^{-1} -\Omega_\mathrm{r} a^{-2} -\Omega_\mathrm{\Lambda} a^2 \\
		& V_\mathrm{{geo}}(a) := -\frac{M}{\Omega_\mathrm{g}-M} \, \left(\frac{3}{4}\Omega_\mathrm{m} a^{-1} + \Omega_\mathrm{r} a^{-2} \right) \\
		& V_\mathrm{{tor}}(a,s) := -\frac{1}{\Omega_\mathrm{g}-M} \, \biggl[ \frac{1}{4}(\dot{s} a + s \dot{a})^2 - s^2 \dot{a}^2 \nonumber\\
		&\quad \quad \quad \quad \quad -\frac{\Omega_s}{2} s^2 a^2  \left(\frac{\dot{a}^2}{a^2}+(\frac{\Omega_s}{2}-1)s^2-\Omega_\mathrm{K} a^{-2}\right)\biggr] + (\Omega_s-1)s^2 a^2 \, .
	\end{align}
\end{subequations}
In contrast to Eq.~\eqref{eq:Fr_ab}, the dot now denotes the derivative with respect to $\tau$ instead of $t$. The equations are complemented by the conservation law \req{eq:theta0} where the initially occurring 3rd derivative of $a$ was replaced with the help of Eq.~\eqref{eq:Fr12b}:
\begin{align}
	\begin{split}
	& - \frac{3}{2}\Omega_\mathrm{m}  a^{-1}\dot{a}\ddot{a} +a^2\ddot{a}s \, [ \, 5\dot{a}s -(2\Omega_s-1) a\dot{s} \, ] \\
	& -(a\dot{s}+\dot{a}s) \left[ \, a^3 \ddot{s}+ 2a^3 s^3 - 5a\dot{a}^2s + 3a^2\dot{a}\dot{s}-4\Omega_\mathrm{g}(\Omega_s-1) a^3 s + 2\Omega_\mathrm{K} a s \, \right] \\  &= 0 \, .
	\label{eq:cl1}
	\end{split}
\end{align}
It should be emphasized that the equations \eqref{eq:Fr12} are not solvable in torsion-free geometry where $s = 0$. This can be seen by taking the time derivative of the first equation and thus eliminating $\ddot{a}$ in the 2nd equation. In this way one obtains \cite{Vasak:2021sce} for $s \equiv 0$:
\begin{equation*}
	\frac{1}{2} \frac{dV}{da} +\frac{V}{a} +2Ma = 0\,.
\end{equation*}
Inserting the potential $V = V_0+V_\mathrm{{geo}}$ gives the obviously wrong relation $ \, 0.75 \,  \Omega_\mathrm{m} a + \Omega_\mathrm{r} = 0$. We therefore conclude that \emph{torsion is necessary for a linear-quadratic ansatz for the free gravity Hamiltonian \eqref{def:Hdyn}}.

Even for the complete system with torsion, however, we face a possible consistency problem, since we are dealing with three equations for the two functions $a(\tau)$ and $s(\tau)$.
Basically, an analogous problem already exists for the Einstein-Friedman equations.
However, the conservation law is automatically satisfied there, since the left-hand side vanishes due to the Bianchi identities and the right-hand side vanishes due to the choice of the equations of state (EOS) for matter and radiation. So far, it has not been possible to provide a proof that the present model is free of contradictions. A recent study based on analytical assumptions \cite{Venn2022} suggests that there is no uniform solution across all 3 cosmological epochs.
However, we will see in the next section that, for selected parameter sets,  numerical solutions exist which also satisfy the conservation law.

\section{Numerical analysis} 
For a numerical analysis of a system of differential equations it is reasonable to put the equations under investigation into the form $y^\prime = f(x,y)$, since numerous proven solution methods are available for this purpose. To achieve this in the present case, the second derivative in the Friedman equation \eqref{eq:Fr12b} has to be removed by introducing a new variable. It seems natural to choose the (dimensionless) Hubble function $H(a) = \dot{a}/a$ with $H(1) = 1$. After some algebra we then get
\begin{subequations} \label{eq:Transf}
	\begin{align}
		\dot{a} & = a \, H  \label{eq:Transfa} \\
		\dot{H} & = -2H^2 +2M +\Omega_{K}a^{-2} -(\Omega_s-1)s^2  \label{eq:Transfb} \\
		\dot{s} & = -Hs + 2\biggl[H^2s^2-M \, \biggl(\frac{3}{4}\Omega_\mathrm{m} a^{-3}+\Omega_\mathrm{r} a^{-4}\biggr) \nonumber \\
		& \quad + \frac{\Omega_s}{2} s^2 \,  \biggl(H^2+(\frac{\Omega_s}{2}-1)s^2-\Omega_{K}a^{-2} \biggr)  \nonumber \\
		& \quad +(\Omega_\mathrm{g}-M) \biggl(H^2+V_0a^{-2}-\Omega_{K}a^{-2}+(\Omega_s-1)s^2 \biggr)\biggr]^{\nicefrac{1}{2}} \enskip . \label{eq:Transfc}
	\end{align}
\end{subequations}
It should be mentioned that, in principle, both signs can appear in front of the root.
The negative sign, however, led in test calculations either to inconsistencies, e.g. violation of the conservation law, or to physically implausible results, e.g. to a growing  scaling factor for $\tau \rightarrow 0$, and is therefore excluded from further analysis here.
The conservation law \eqref{eq:cl1}, expressed with the variables $a,H,s$, now reads
\begin{align} \label{eq:cl2}
	\begin{split}
		&\biggl[\frac{3}{2}\Omega_\mathrm{m} a^{-3} H- s \,  \biggl(5Hs-(2\Omega_s-1)\dot{s}\biggr)\biggr] \biggl[H^2-\Omega_\mathrm{K}a^{-2}-2M +(\Omega_s-1)s^2\biggr] \\
		& -(Hs+\dot{s}) \, 	\biggl[\ddot{s}+2s^3-5H^2s+3H\dot{s}-4\Omega_\mathrm{g}(\Omega_s-1)s+2\Omega_\mathrm{K}a^{-2}s\biggr]
		= \, 0 \, .
	\end{split}
\end{align}
These equations have to be solved with suitable boundary conditions.
Without loss of generality we set the variable $a(\tau_1) = 1$ for $\tau_1=1$, the present time. Since the Hubble function $H$ has already been "normalized" to $H_{0}$, its present value, $H(1) = 1$ is automatically valid.
However, there is no obvious choice for the initial value of the torsion parameter $s_1 \equiv s(1)$.
Even the relation for the cosmological parameters derived from the first Friedman equation does not provide any remedy.
To show this, we replace the potential $V$ in the first Friedman equation with the densities $\hat{\rho}_i$ defined relative to the critical density \eqref{eq:rho_cr}:
\begin{align} \label{eq:H2}
	\begin{split}
		H^2  = \frac{\dot{a}^2}{a^2} = \frac{1}{a^2}(-V+\Omega_\mathrm{K}) = \hat{\rho}_\mathrm{m}+\hat{\rho}_\mathrm{r}+\hat{\rho}_\mathrm{\Lambda}+\hat{\rho}_\mathrm{K}+\hat{\rho}_\mathrm{{geo}}+\hat{\rho}_\mathrm{{tor}}
	\end{split}
\end{align}
with \footnote{Interpreting all terms on the r.h.s. of Eq.~\eqref{eq:H2} as relative energy densities,  we recover the zero-energy condition of Eq.~\eqref{eq:modEinstein} in the form
	\begin{equation}
		\rho_\mathrm{st} + \rho_\mathrm{matter} = 0 \, . \nonumber
	\end{equation}
	Hereby the relative Hubble parameter that depends on the expansion velocity of the Universe is naturally identified with the relative kinetic energy of spacetime while the other geometry-related energy densities play the role of potential energy densities:
	\begin{equation}
		\rho_\mathrm{st} := -H^2(\dot{a},a) +\rho_\mathrm{r}+\rho_\mathrm{\Lambda}+\rho_\mathrm{K}+\rho_\mathrm{{geo}}+\rho_\mathrm{{tor}}\, . \nonumber
	\end{equation}
	Obviously $\rho_\mathrm{st} \le 0$ for $\rho_\mathrm{matter} \ge 0$ -- a ghost term that together with the structural elements of the geometry absorbs the energy density of matter.}
\begin{align}
	\begin{split}
		& \hat{\rho}_\mathrm{m} = \Omega_\mathrm{m} \, a^{-3} \\
		& \hat{\rho}_\mathrm{r} = \Omega_\mathrm{r} \, a^{-4} \\
		& \hat{\rho}_\mathrm{\Lambda} = \Omega_\mathrm{\Lambda} \\
		& \hat{\rho}_\mathrm{K} = \Omega_\mathrm{K} \, a^{-2} \\
		& \hat{\rho}_\mathrm{{geo}} = \Omega_\mathrm{{geo}} (a) := -V_\mathrm{{geo}} \, a^{-2} \\
		& \hat{\rho}_\mathrm{{tor}} = \Omega_\mathrm{{tor}} (a,s) := -V_\mathrm{{tor}} \, a^{-2} \, .
	\end{split}
\end{align}

We mark today's values with index 1 and obtain
\begin{align}
	\begin{split}
		& \Omega_{\mathrm{geo},1} = \frac{(\frac{1}{4}\Omega_\mathrm{m}+\Omega_\mathrm{\Lambda})\,(\frac{3}{4}\Omega_\mathrm{m}+\Omega_\mathrm{r})}{\Omega_\mathrm{g} - \frac{1}{4}\Omega_\mathrm{m}-\Omega_\mathrm{\Lambda}} \\
		& \Omega_{\mathrm{tor},1} = \frac{\frac{1}{4}(\dot{s}_1+s_1)^2 -s_1^2 -\frac{\Omega_s}{2}\, s^2_1\,(1+(\frac{\Omega_s}{2}-1)s^2_1 -\Omega_\mathrm{K})}{\Omega_\mathrm{g} - \frac{1}{4}\Omega_\mathrm{m}-\Omega_\mathrm{\Lambda}} -(\Omega_s-1)\,s_1^2
		\label{eq:geo_s} \, ,
	\end{split}
\end{align}
which yields a relation containing besides $s_1$ also its unknown first derivative $\dot{s}_1$:
\begin{align}
	1 = \Omega_\mathrm{m}+\Omega_\mathrm{r}+\Omega_\mathrm{\Lambda}+\Omega_\mathrm{K}+\Omega_{\mathrm{geo},1}+\Omega_{\mathrm{tor},1} \, .
\end{align}
Compared with the standard parameter set of the Concordance model, $\Omega_\mathrm{m}, \, \Omega_\mathrm{r}, \, \Omega_\mathrm{\Lambda}$ and $\Omega_\mathrm{K}$, there are three new independent parameters in this theory, namely $g_1$ in $\Omega_\mathrm{g}$, $g_3$ in $\Omega_s$, and the initial value $s_1$ for which no specific observational data are available and whose range of values cannot be limited a priori - except that $\Omega_\mathrm{g}$ must be non-zero, otherwise $g_1$ becomes infinite, and the root in \req{eq:Transfc} must be real. In the next section we will carry out numerical tests based on the parameter sets as listed in~\rtab{tab:defparam}.

However, upon assuming the zero-energy condition for the Universe, that set gets reduced by one parameter. It is \cite{Vasak:2022gps} that $\Omega_\mathrm{\Lambda} = 0$ introduces the dependence \eqref{eq:gi(g4)} of $g_1$ and $g_2$ on the vacuum energy density of matter, $g_4$, which then replaces $\Omega_\mathrm{\Lambda}$. This will lead to a substitution of the dark energy role of the cosmological constant by the torsion density of spacetime.

\section{First Results} \label{sec:result}
We start the numerical calculation applying a 4th order Runge-Kutta method with step size adjustment, and set arbitrarily $s_1=0.5$. To achieve comparability of the torsion terms in \req{eq:consist} with the Einstein tensor, we set $2g_3 =1/8 \pi G$ from which $\Omega_\mathrm{s} = 0.5$ follows according to the definition \eqref{eq:def_Os} . Furthermore we use for $\Omega_\mathrm{m} h^2 = (\Omega_\mathrm{b} + \Omega_\mathrm{c}) h^2$, $\Omega_\mathrm{\Lambda}$ and $h$ the cosmological parameters from \textit{Planck} TT,TE,EE+lowE+lensing, \cite{planck15}, set $\Omega_\mathrm{r} h^2 = 2.47 \cdot 10^{-5}$ corresponding to the temperature $T = \SI{2.725}{\kelvin}$ of CMB,  and $\Omega_\mathrm{K} = 0$ assuming a flat Universe, see \rtab{tab:defparam}, 1st line. Note that the value of $\Omega_\mathrm{\Lambda}$ is not measured but derived following the base-$\Lambda$CDM cosmology, according to which $\Omega_\mathrm{\Lambda} = 1 -\Omega_\mathrm{m}-\Omega_\mathrm{r}$ holds.
\begin{table}[ht]
	\centering
	\begin{tabular}[t]{c   c   c   c   c}
		\toprule
		$\quad h \quad$ & $ \quad \Omega_\mathrm{m}  \quad $ & $ \quad \Omega_\mathrm{r} \quad $ & $ \quad \Omega_\mathrm{\Lambda} \quad $ & $ \quad \Omega_\mathrm{K} \quad $ \\
		\midrule
		$0.674$ & $0.315$ & $5.44\cdot 10^{-5}$ & $0.685$ & $0.000$ \\
		$0.740$ & $0.261$ & $4.51\cdot 10^{-5}$ & $0.000$ & $0.000$ \\
		\botrule
	\end{tabular}
	\caption{\footnotesize The $\Lambda$CDM parameter set $h, \, \Omega_i$: 1st row from Planck Collaboration \cite{planck15}, Table 2, column 5.  2nd row obtained by adjusting the original \textit{Planck} values in relation to the changed $h$.}
	\label{tab:defparam}
\end{table}

\begin{figure}[ht]
	\centering
	\begin{tabular}{cc}
		\includegraphics[scale=0.29]{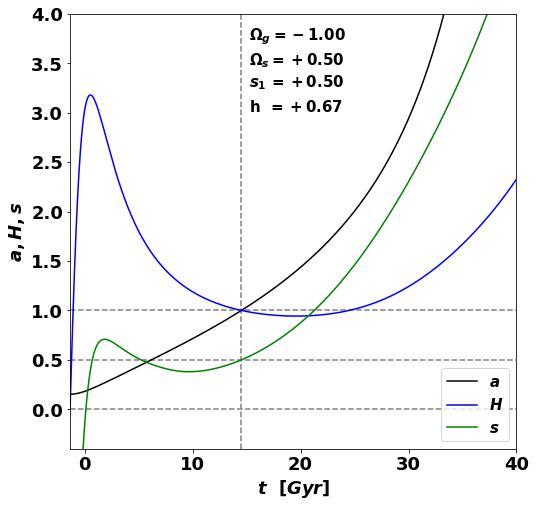} &
		\includegraphics[scale=0.29]{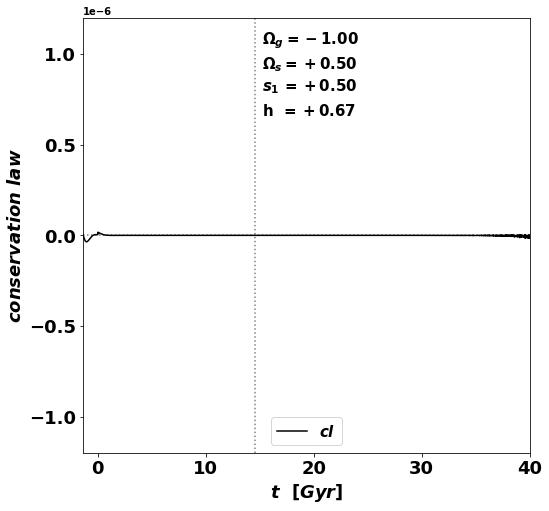}
	\end{tabular}
	\caption{\footnotesize Left panel: $a$, $H$, and $s$ as a function of $t$, where the scaling factor $a$ reaches its minimum for a time less than zero and increases strictly monotonically thereafter. Right panel: Prove that the conservation law $cl(t)=0$ is fulfilled (note the scaling of the ordinate). The small deviations from the zero line are exclusively of numerical origin and can be significantly reduced by suitable selection of the step size. The vertical dashed line marks the Hubble time.}
	\label{fig:aHs+law}
\end{figure}

An example depicted in~\rfig{fig:aHs+law} shows the result for $\Omega_\mathrm{g} = -1$.
We observe that the system of equations \eqref{eq:Transf} is completely solvable (left panel) and consistent in the given domain (right panel). From a physical point of view, it is noteworthy that the scaling factor $a$ is significantly different from zero at the origin $t=0$ and even for negative times. Times less than zero are no objection against our theory, in which the present time was arbitrarily identified with the Hubble time $t = 1/H_0$. However, this agrees only with the age of the Universe in the case of a uniform expansion.

\begin{figure}[ht]
	\centering
	\begin{tabular}{cc}
		\includegraphics[scale=0.28]{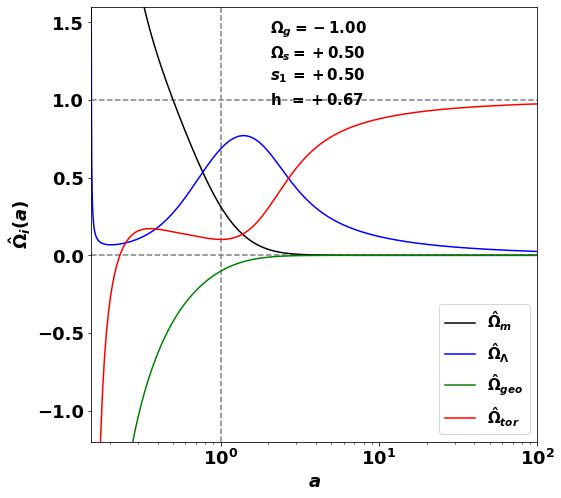} &
		\includegraphics[scale=0.28]{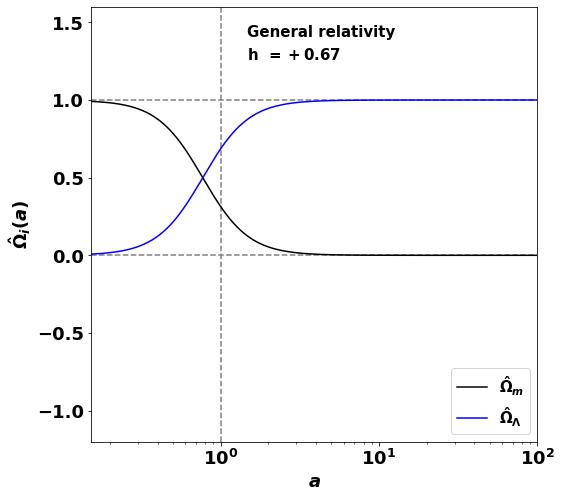}
	\end{tabular}
	\caption{\footnotesize Evolution of the fractional density parameters $\hat{\Omega}_i(a) = \hat{\rho}_i(a)/H^2(a)$ as a function of $a$. For the sake of clarity, the parameters $\hat{\Omega}_\mathrm{r}$ and $\hat{\Omega}_\mathrm{K}$ have been neglected as they do not contribute in the given domain. Note that the representation of $\hat{\Omega}_\mathrm{{tor}}(a(t),s(t))$ as a function of $a$ (left panel) is only feasible because $a(t)$ grows strictly monotonically with time. Right panel: $\hat{\Omega}_\mathrm{m}$ and $\hat{\Omega}_\mathrm{\Lambda}$ of general relativity.}
	\label{fig:oga}
\end{figure}

Another important physical aspect is revealed in~\rfig{fig:oga} showing (left panel) the evolution of the fractional density parameters defined by $ \, \hat{\Omega}_i (a) := \hat{\rho}_i(a) / H^2(a)$  for $ \, i = \mathrm{m,\Lambda,geo,tor}$. For large scale factors $a$, the torsion term $\hat{\Omega}_\mathrm{{tor}}$  dominates all others, in particular $\hat{\Omega}_\mathrm{\Lambda}$ is diminishing. In contrast, $\hat{\Omega}_\mathrm{\Lambda}$ dominates in the conventional FLRW approach.

\section{Torsion for Dark Energy} \label{sec:torsionalDE}
\begin{figure}[ht]
	\centering
	\begin{tabular}{cc}
		\includegraphics[scale=0.29]{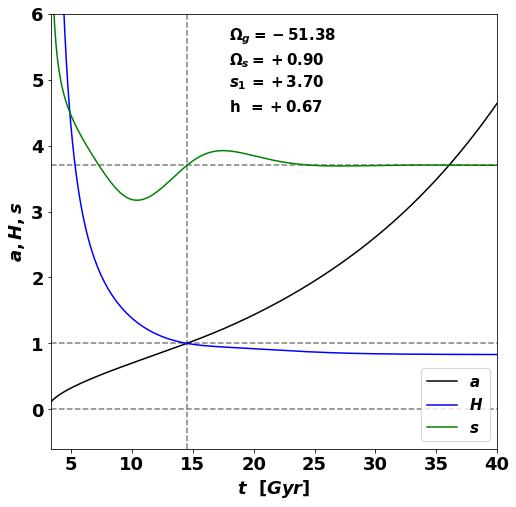}  & \includegraphics[scale=0.29]{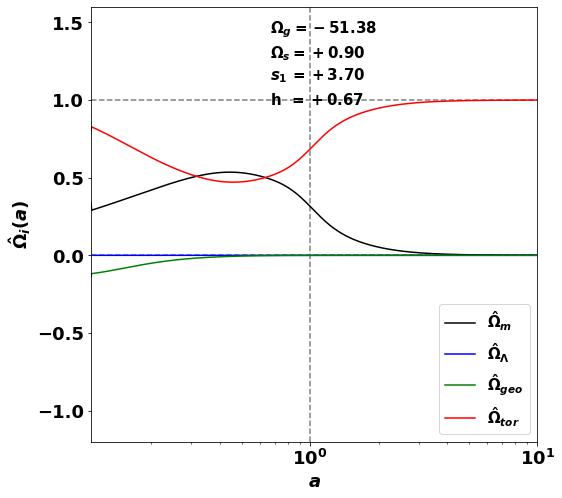} \\
		\includegraphics[scale=0.29]{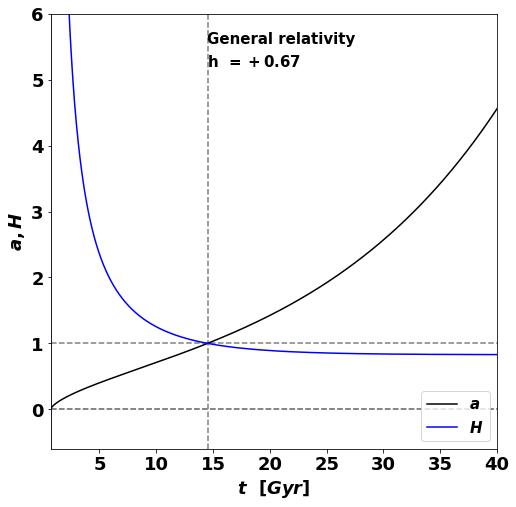} &
		\includegraphics[scale=0.29]{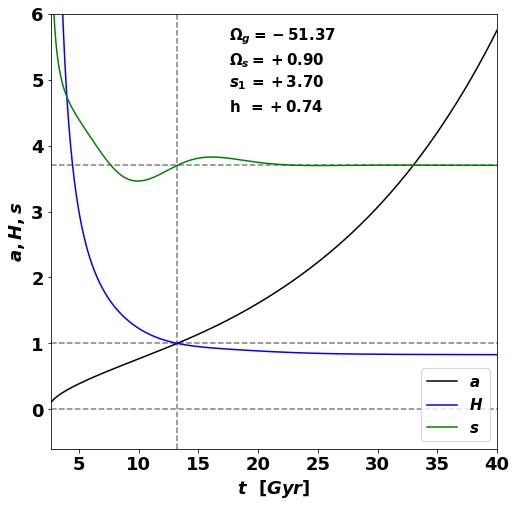}
	\end{tabular}
	\caption{\footnotesize Upper left panel: The scaling factor a (shown in black) starts at a positive value and develops exponentially at times $t \sim 1/H_0$ and later. Upper right panel: The fractional torsion density $\hat{\Omega}_\mathrm{{tor}} (a)$ in close agreement with the classical approach for $a \geq 1$ (red curve), see \rfig{fig:oga}, right panel. As expected, the fractional matter density $\hat{\Omega}_\mathrm{m} (a)$ fades away in an expanding Universe (black curve). The same is true for the fractional density $\hat{\Omega}_\mathrm{{geo}} (a)$ of a deformed spacetime, which flattens out with increasing expansion (green curve). According to our preposition, there is no contribution of the cosmological constant (blue curve). The lower left panel has been calculated for a base-$\Lambda$CDM Universe with the same \textit{Planck} parameter set, on the right the result for an increased $h$ and $\Omega $-parameters after \rtab{tab:defparam}, 2nd line.}
	\label{fig:nolambda}
\end{figure}

We thus follow the zero-energy condition derived in \cite{Vasak:2022gps} and set $\Lambda_0 = 0$ respectively $\Omega_\Lambda = 0$ in order to investigate whether torsion can partially or even fully explain all the phenomena attributed to dark energy facilitated otherwise by the cosmological constant. With this assumption it is easy to show that the equations \eqref{eq:Transf}
can be solved exactly in the asymptotic domain $\tau \rightarrow \infty$. From
\begin{align}
	a(\tau \rightarrow \infty) = e^{H_\mathrm{\infty} \tau}
\end{align}
with an asymptotically constant Hubble function $H_\mathrm{\infty}$ follows $\dot{H} =0$ and as per \req{eq:Transfb}
\begin{align}
	s^2 _\mathrm{\infty} = \frac{2 H^2 _\mathrm{\infty}}{1-\Omega_\mathrm{s}}. \label{eq:s_inf}
\end{align}
\req{eq:Transfc} finally leads to
\begin{align}
	H^2 _\infty = 2 \, \Omega_\mathrm{g} \frac{(1-\Omega_\mathrm{s})^2}{3-5 \, \Omega_\mathrm{s}}
\end{align}
This enables us to choose the still free parameter $\Omega_\mathrm{g}$ in such a way that the Hubble function has the same asymtotic as in the standard theory, that is $H_\infty = H_\mathrm{GR,\infty} = \sqrt{\Omega_\mathrm{GR,\Lambda}}$:
\begin{align}
	\Omega_\mathrm{g} = \frac{1}{2} \, H_\mathrm{GR,\infty}^2 \, \frac{3-5 \, \Omega_\mathrm{s}}{(1-\Omega_\mathrm{s})^2} \, .
	\label{eq:Og_inf}
\end{align}
In addition, we assume $s_1 = s_\mathrm{\infty}$ which means
\begin{align}
	s_1 = H_\mathrm{GR,\infty} \, \left(\frac{2}{1-\Omega_\mathrm{s}}\right)^{\nicefrac{1}{2}} \, .
\end{align}
The only remaining free parameter $\Omega_\mathrm{s}$ must be less than 1 by \req{eq:s_inf} and greater than 0.6 by \req{eq:Og_inf} to allow AdS geometry, i.e. $\Omega_\mathrm{g} \, < \, 1$ respectively $g_1 \, < \, 0$, as discussed in Ref. \cite{Vasak:2022gps}. Thus we set for the following example calculation $\Omega_\mathrm{s} = 0.9$. As we can see in~\fref{fig:nolambda}, upper left panel, there is an exponential progression of $a(t)$ as well as a dominance of the torsion-related fractional density for future times (upper right panel). Most interesting is the excellent agreement with the result for the base-$\Lambda$CDM Universe (lower left panel), based on the same parameters $h$, $\Omega_\mathrm{m}$, and $\Omega_\mathrm{r}$, however with a non vanishing $\Omega_\mathrm{GR,\Lambda} \, = \, 0.685$.

We therefore conclude: The presented results suggest that torsion is well suited to play the role of dark energy. Even more, the relation $\hat{\Omega}_{\mathrm{tor},1}/\hat{\Omega}_\mathrm{m} \sim 0.75 / 0.25$ observed in~\fref{fig:nolambda} sheds new light on the so called ``Coincidence Problem''!

Another interesting insight is found in Ref. \cite{Venn2022}. It becomes visible when we resolve \req{eq:Og_inf} to $\Omega_\mathrm{s}$:
\begin{align}
	\Omega_\mathrm{s, \pm} = -\frac{5}{4} \, \frac{H_\mathrm{GR,\infty}^2}{\Omega_\mathrm{g}} +1 \pm \left[\left(\frac{5}{4} \frac{H_\mathrm{GR,\infty}^2}{\Omega_\mathrm{g}}\right)^2-  \frac{H_\mathrm{GR,\infty}^2}{\Omega_\mathrm{g}}\right]^{\nicefrac{1}{2}} \, .
\end{align}
Real values of $\Omega_\mathrm{s}$ are obtained if
\begin{align}
	\Omega_\mathrm{g} \le \frac{25}{16} \, H_\mathrm{GR,\infty}^2 \, .
\end{align}
We now replace $\Omega_\mathrm{g}$ by its definition \eqref{eq:def_Og} and use the relation \eqref{eq:gi(g4)} between the deformation parameter $g_1$ and the the vacuum energy of matter $g_4$ (which only holds for $\Omega_\mathrm{\Lambda} = 0$) to get
\begin{align}
	g_4 \ge -\frac{75}{8} M_p^2 \, H_0^2 \, H_\mathrm{GR,\infty}^2 \, .
\end{align}
This results in a lower bound of $ \simeq -8.5 \, e^{-47} \, \mathrm{GeV^4} \simeq -2 \, e^{-29} \, \mathrm{\nicefrac{g}{cm^3}}$ for the vacuum energy.
Finally we evaluate the impact of $h$ and set $h = 0.740$ according to the local measurements of Riess et al. \cite{Riess2019b} observing the Cepheids in the Large Magellanic Cloud. The associated values for $\Omega_i$ are gained by simply scaling $\Omega_i(\text{Riess}) = \Omega_i(\text{Planck}) \cdot  (h_{\text{Planck}}/h_{\text{Riess}})^2, \, i= \mathrm{m,r}$ and depicted in \rtab{tab:defparam}, 2nd line.
The lower right panel of \rfig{fig:nolambda} shows the result: The exponential growth of $a(t)$ is steeper just as expected for larger $h$ and the Hubble time is correspondingly shorter. But the qualitative course does not change.

\section{Conclusion}\label{sec:ccl}
The derivation of a covariant gauge theory of gravity from fundamental principles and its application to cosmology lead to an extension of the Friedman-Lemaître equations of the $\Lambda$CDM standard model.
However, with a linear-quadratic approach of the Hamiltonian for the dynamical gravitational field, the corresponding FLRW cosmology is consistent only if torsion is taken into account and the following two conditions are satisfied: Firstly, the covariant conservation of the stress-energy tensor must be ensured. That is accomplished by modelling torsion with a completely anti-symmetric tensor.
Secondly, the cosmological principle must be preserved, which is achieved by substituting that torsion tensor with a time-like homogeneous axial vector $(s_0(t),0,0,0)$. The equations for the scaling factor $a(t)$ and the torsion function $s(t)$ obtained in this way can be solved for selected parameter sets while simultaneously satisfying the covariant conservation law.

Taking the parameters of the standard model as given, three free parameters remain in this concept: The deformation parameter $\Omega_\mathrm{g}$, the torsion parameter $\Omega_\mathrm{s}$, and the value of $s$ at the present time.
The result for the more or less arbitrarily chosen set $(\Omega_\mathrm{g}, \Omega_\mathrm{s}, s_1) = (-1, 0.5, 0.5)$ led to the conjecture that torsion can at least contribute to dark energy. Via an asymptotic consideration it could be shown that the expansion of the Universe can be described with the help of the torsion, and without any contribution of the cosmological constant, equivalently with the base-$\Lambda$CDM model.

Although the above results do not represent a definite confirmation, they nevertheless offer a reasonable \emph{indication} that torsion can resolve both mysteries, the magnitude and the coincidence problems, ascribed to the cosmological constant, a quantity which is the subject of much speculation in modern physics. A comprehensive comparison with observational data is needed, and work to apply a full fledged MCMC analysis is in preparation.

\backmatter

\bmhead{Acknowledgments}
The authors are indebted to the ``Walter Greiner-Gesellschaft zur F\"{o}rderung der physikalischen
Grundlagenforschung e.V'' (WGG) in Frankfurt for their support. JK, DV and AV especially thank the Fueck Stiftung for support. The authors also wish to thank David Benisty for valuable discussions.

\input torsion.bbl


\end{document}